\documentclass{epl}
\usepackage[english]{babel}
\usepackage[T1]{fontenc}
\usepackage{amsmath}
\usepackage{graphicx}

\newlength{\smallmathskip}
\newlength{\medmathskip}
\newlength{\bigmathskip}
\newcommand{\parz}[2]{\frac{\partial #1}{\partial #2}}
\newcommand{\tdiff}{\operatorname{d}\!}
\newcommand{\Lop}{\operatorname{\mathcal{L}}}
\newcommand{\varexp}{\operatorname{e}}

\title{\centerline{Macroscopic Aharonov--Bohm Effect in Type-I Superconductors}}

\author{Catrin Ellenberger, Florian Gebhard, Wolfgang Bestgen}

\shortauthor{Catrin Ellenberger et al.} 
\shorttitle{Macroscopic Aharonov--Bohm Effect} 

\institute{Fachbereich Physik, Philipps--Universit\"{a}t Marburg,  
D--35032 Marburg, Germany}


\begin{document}

\maketitle 
\date{ \today; Europhys.~Lett.~{\bf 69}, 431 (2005).}

\begin{abstract}%
In type-I superconducting cylinders bulk superconductivity is destroyed
above the first critical current. Below the second critical current 
the `type-I mixed state' displays fluctuation superconductivity
which contributes to the total current. A magnetic flux on the
axis of the cylinder can change the second critical current
by as much as 50~percent so that half a flux quantum can switch
the cylinder from normal conduction to superconductivity:
the Aharonov--Bohm effect manifests itself in 
macroscopically large resistance changes of the cylinder.
\end{abstract} 

\section{Introduction}

The superconducting condensate can be used
as a `quantum amplifier'. For example, microscopic effects such as 
quantum interference of electron waves become macroscopically measurable
as flux quantization. In this work the macroscopic condensate
is used to make the Aharonov--Bohm effect prominent
on macroscopic scales.

Consider a wire of radius~$R$ with a current~$J$ parallel to the
wire axis. The superconductivity of a type-I superconductor breaks down
when the current exceeds the Silsbee current $J_{\text{c1}}$.
The Silsbee current is determined from the condition that
the magnetic field at the surface of the wire is equal to
the upper critical field $H_{\text{c}}$,
$J_{\text{c1}}=c_0 R H_{\text{c}}/2$, where $c_0$ is the speed of light.
For $J>J_{\text{c1}}$, the so-called intermediate state forms in
the wire with alternating normal and superconducting domains~\cite{london}.
The diameter of the superconducting domains~$d$ shrinks and finally
the intermediate state collapses when $d$~becomes of the order of the
coherence length~$\xi$. However, superconductivity still remains in
the form of small temporal fluctuations. 
In this `mixed state' for type-I superconductors,
superconducting fluctuations and electrical and magnetic fields co-exist.
The mixed state was first predicted
by L.D.~Landau in 1938~\cite{landau} as a surface state for hollow superconducting 
cylinders. Its existence  was later verified experimentally
by I.L.~Landau and Yu.V.~Sharvin~\cite{landau69}.
As later shown by Andreev and Bestgen~\cite{andreevbestgen} the fluctuation induced
superconductivity is destroyed when the thickness
of the fluctuation layer~ becomes smaller 
than the coherence length~$\xi$, and superconductivity is
ultimately destroyed at the second critical current~$J_{\text{c2}}$
which was first calculated by~Andreev~\cite{andreev}.
An externally applied magnetic field~$H_{\text{a}}$ parallel to
the wire axis contributes
to the suppression of the superconducting phase and reduces
$J_{\text{c1}}$ and $J_{\text{c2}}$ accordingly~\cite{bestgen79,bestgen80}. 

In this work we investigate the stability of the mixed state
in type-I superconductors for a hollow cylinder of inner 
radius~$R_{\text{i}}\ll R$
in a magnetic field~$H_{\text{a}}$ with an additional magnetic flux line 
of strength~$\Phi_{\text{a}}$ on the cylinder axis. 
We show that the Aharonov--Bohm effect leads to strong oscillations
of the second critical current~$J_{\text{c2}}(\Phi_{\text{a}})$ 
as a function of $\Phi_{\text{a}}/\Phi_0$
where $\Phi_0=hc_0/(2e)$ is the flux quantum. 
The effect is particularly marked for the 
`one-dimensional' mixed state for which $R_{\text{i}}\ll \xi$.
In this case, $J_{\text{c2}}(\Phi_{\text{a}})$ can vary by almost 50~percent.
If the applied current through the wire is fixed at some value
$J_{\text{c1}} < J_{\text{c2}}(\Phi_0/2) < J < J_{\text{c2}}(0)$,
the modulation of~$\Phi$ between integer flux quanta will lead
to a periodic onset and breakdown of the superconducting 
current~$J_{\text{s}}$ as a function of~$\Phi$. 
The resistance~$\Omega$ at applied voltage~$V$ becomes
\begin{equation}
\Omega = \frac{V}{J}=\frac{V}{J_{\text{n}} +J_{\text{s}}^z}\approx \Omega_{\text{n}}
\left( 1- \frac{J_{\text{s}}^z}{J}\right)\; .
\label{ohm}
\end{equation}
Therefore, the Aharonov--Bohm effect leads to periodic fluctuations 
in the resistance
of type-I superconductors in the mixed state.

\section{Model and superconducting current}

In the mixed state with $J\simeq J_{\text{c2}}$,
the dynamics of the superconducting fluctuations at the inner cylinder
surface is described by the time-dependent, linear Ginzburg--Landau
equations with Langevin forces. In dimensionless cylindrical
coordinates $\rho,\phi,z$ they read~\cite{bestgen80,cediss,bestgen74}
\begin{equation}
  \parz{\Psi}{t}-\Psi +\Lop{\Psi} = f(\rho,\varphi,z,t)\;\text{,}
\end{equation}
with the differential operator
\begin{equation}
  \Lop =-\frac{1}{\rho}\parz{}{\rho}\left(\rho \parz{}{\rho}\right)-\left(\frac{1}{\rho}\parz{}{\rho}-\frac{i\gamma}{2}\rho-\frac{i\delta}{\rho}\right)^2
  -\left(\parz{}{z}+\frac{i\beta}{2}\rho^{2}+i\epsilon t\right)^2
\end{equation}
and the Gaussian fluctuations
\begin{equation}
  \left\langle f(\rho,\varphi,z,t)\:f^{\ast}(\rho',\varphi',z',t')\right\rangle=  \frac{1}{\rho}\delta\!\left(\rho-\rho'\right)\delta\!\left(\varphi-\varphi'\right)\delta\!\left(z-z'\right)\delta\!\left(t-t'\right)
\end{equation}
The dimensionless parameters $\epsilon$, $\beta$, $\gamma$ and $\delta$
are given by
\begin{alignat}{2}
  \epsilon&=\frac{2 e \xi}{\hbar\nu}E\;\text{,} &\qquad \gamma&=
\frac{2 e H_{\text{a}}\xi^{2}}{\hbar c_{0}}=\frac{H_{\text{a}}}{H_{\text{c2}}}
\;\text{,}\notag\\
  \beta&=\frac{4 e \xi^{3} J}{\hbar \text{c0}^{2} R^{2}}\;\text{,} &\qquad \delta&=\frac{e \Phi_{\text{a}}}{\hbar c_{0} \pi}=\frac{\Phi_{\text{a}}}{\Phi_{0}}\;\text{.}
\end{alignat}
$\xi=\xi(T)$ is the coherence length, 
$\nu=8k_{\text{B}}(T_{\text{c}}-T)/(\pi\hbar)$
is the relaxation frequency, and $e$ and $m$ are the electron charge and mass,
respectively. $E$ is the electrical field which drives the total
current~$J=J_{\text{n}}+J_{\text{s}}^z$ in $z$-direction.
For type-I superconductors was shown in~\cite{andreevbestgen} that $\epsilon\ll 1$. 
Using the linearized Ginzburg--Landau equations implies
that $J_{\text{s}}^z \ll J$. 

The general expression for the total longitudinal fluctuation
current is calculated along the lines of~\cite{andreevbestgen,bestgen74} and gives the result
\begin{equation}
  J^{\text{s}}_{z}=\frac{2 e T k_{\text{B}}}{\pi\epsilon\hbar} \sum_{n m}\int\limits_{-\infty}^{\infty}\tdiff k\,\parz{\lambda_{n m k}}{k} F_{n m}(k)\;\text{,}
\end{equation}
with
\begin{equation}
  F_{n m}(k)=\int\limits_{0}^{\infty}\tdiff\tau\,\exp\!\left[\frac{2}{\epsilon}\left(\tau- \int\limits_{k-\tau}^k \lambda_{n m k'}\,\tdiff k'\right)\right]\;\text{.}
\end{equation}
$\lambda_{n m k}$ is the lowest eigenvalue of $\Lop$ as a function of $k=k_{z}+\epsilon t$ where $k_{z}$ is the wavenumber vector in $z$-direction. The spectrum $\lambda_{n m k}$ can be obtained with the help of a variational method, which starts from the ansatz $R_{n=0,m k}\sim\rho^{|m|}\varexp^{-\alpha\rho^{2}}$ for the radial part of $\psi$.
We may restrict ourselves to the dominant contribution, $n=0$~\cite{bestgenrothen};
$\alpha$ is a variational parameter which determines the radial extension
of the superconducting region\cite{andreevbestgen,bestgen74}. From the variation principle we then find
for $\lambda_{m,k_m}=\lambda_{0,m,k}^{\text{opt}}$
\begin{equation}
  \lambda_{m k_{m}}=-\left(m-\delta\right)\gamma+4\alpha\left(|m-\delta|+1\right)
  -\frac{\beta^{2}}{16\alpha^{2}}\left(|m-\delta|+ 1\right),
\end{equation}
with
\begin{gather}
  2-\frac{\gamma^{2}+4 k_{m}\beta}{8\alpha^{2}}-\frac{\beta^{2}\left(|m-\delta|+2\right)}{8\alpha^{3}}=0\;\text{,}\\[\smallmathskip]
  k_{m}=\frac{-\beta\left(|m-\delta|+1\right)}{4\alpha}\;\text{.}
\end{gather}

The critical current can be evaluated analytically in the vicinity
of the critical curve~\cite{cediss} but for our further analysis explicit 
expressions for $J_{\text{s}}^z$ are not required. We merely state
that the experimental parameters can be tuned in such a way that
the fluctuation current is several percent of the total current,
i.e., the resistance fluctuations according to~(\ref{ohm})
should be measurable.

\section{Second critical current}

In our theory, the critical current $J_{\text{c2}}(H_{\text{a}},\Phi_{\text{a}})$
above which the superconducting fluctuations vanish is determined
by the condition
\begin{equation}
\mathop{\text{Min}}_{m}\left[ \lambda_{m,k_m}(\beta,\gamma,\delta)\right]=1
\label{(8)} \; .
\end{equation}
If the minimum over all $\lambda_{m,k_m}$ was noticeably smaller than unity
the corresponding supercurrent would become so large that the use
of a linearized Ginzburg--Landau equations would not have been justified.
If the minimum over all $\lambda_{m,k_m}$ was substantially larger than unity
$J_{\text{s}}^z$ would be tiny. Our theory
applies to the transition region where the supercurrent is small
but measurable.

\begin{figure}[htbp]
\begin{center}
\includegraphics[width=14cm]{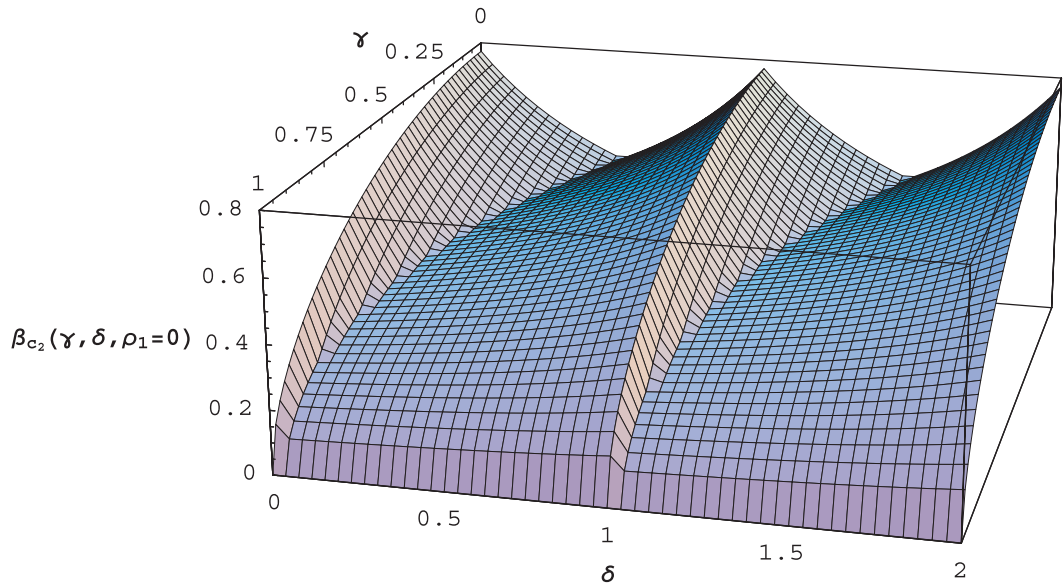}
\end{center}
\caption{The normalized second 
critical current $\beta_{\text{c2}}\left(\gamma, \delta\right)$
for $R_{\text{i}}\ll \xi$.\label{fig1}}
\end{figure}

Solving~(\ref{(8)}) for $J_{\text{c2}}$ leads to 
the second critical current as a function of the external magnetic field
and applied flux in the interior of the cylinder. In SI~units we find:
\begin{eqnarray}
J_{\text{c2}}\left(\frac{H_{\text{a}}}{H_{\text{c2}}}
\frac{\Phi}{\Phi_{0}}\right)
&=&\beta_{\text{c2}} \frac{c_{0}^{2}R^{2}\hbar}{4e\xi^{3}}\nonumber\\
&=&\frac{c_{0}^{2}R^{2}\hbar}{e\xi^{3}}
\Biggl[A^{3}
\left(1+\sqrt{1-
\left(\frac{H_{\text{a}}}{H_{\text{c2}}}\right)^{2}
\frac{1}{48A^{2}}}
\right)^{3}\\
  && -\left(\frac{H_{\text{a}}}{H_{\text{c2}}}\right)^{2}
\frac{A}{16}
\left(1+\sqrt{1-\left(\frac{H_{\text{a}}}{H_{\text{c2}}}\right)^{2}
\frac{1}{48A^{2}}}\right)\Biggr]^{1/2}\;\text{,}\nonumber
\label{(9)}
\end{eqnarray}
where
\begin{equation}
  A=\frac{1+\left(m-\delta\right)\gamma}{6\left(|m-\delta|+1\right)}\;\text{.}
\end{equation}
For a given flux~$\delta$, $m$ in~(\ref{(9)}) is the integer which
gives the minimal~$J_{\text{c2}}$\cite{cediss}. Figure~\ref{fig1} 
gives a three-dimensional plot of $J_{\text{c2}}$ as a function of $\gamma$ (external field) and $\delta$ (central flux).

The second critical current can be 
almost halved by the application 
of a suitable vector potential 
even if no magnetic field is present. 
This is a macroscopic realization of the Aharonov-Bohm effect.

\section{Conclusions}

In this work we have studied the supercurrent in a hollow cylinder
of type-I superconducting material in the presence of an external
magnetic field and a flux line on the cylinder axis. 
We have shown that the second critical current, above which
fluctuation superconductivity breaks down, periodically changes
as a function of the applied flux, by as much as 50~percent 
in the absence of an external field. These Aharonov--Bohm oscillations
lead to a periodic change of the resistance of the cylinder, see~(\ref{ohm}).
The latter are measurable because the currents due to the 
superconducting fluctuations are sizable, of the order of a few percent
of the total current. Therefore, a suitably tuned experiment
with type-I superconducting cylinders 
should be able to verify the Aharonov--Bohm effect
by macroscopic resistance measurement.

\acknowledgments
We thank F.~Rothen and M.~Paech for useful discussions.


\begin{thebibliography}{99}
\bibitem{london}
\Name{London F.} 
\Book{Superfluids I.} 
\Publ{Wiley, New York}
\Year{1950}.
\bibitem{landau} 
\Name{Landau L.D.} 
private communication with D.\ Shoenberg; see: D. Shoenberg \textsl{Superconductivity.} Cambridge: Cambridge University Press, 1938; page 59.
\bibitem{landau69} 
\Name{Landau I.L.\and Sharvin Yu.V.} 
\REVIEW{Pis'ma Zh.\ Eksp.\ Teor.\ Fiz.}{10}{1969}{192} 
 and 
\REVIEW{JETP Lett.}{10}{1969}{121}.
\bibitem{andreevbestgen} 
\Name{Andreev A.F.\and Bestgen W.} 
\REVIEW{Zh.\ Eksp.\ Teor.\ Fiz.}{10}{1973}{453} and 
\REVIEW{Sov.\ Phys.\ JETP}{37}{1973}{942}.
\bibitem{andreev} 
\Name{Andreev A.F.} 
\REVIEW{Pis'ma Zh.\ Eksp.\ Teor.\ Fiz.}{ 10}{1969}{453} and \REVIEW{JETP Lett.}{10}{1969}{291}.
\bibitem{bestgen79} 
\Name{Bestgen W.} 
\REVIEW{Zh.\ Eksp.\ Teor.\ Fiz.}{76}{1979}{566} and 
\REVIEW{Sov.\ Phys.\ JETP}{49}{1979}{285}.
\bibitem{bestgen80} 
\Name{Bestgen W.} 
\REVIEW{Sol.\ State Commun.}{36}{1980}{441}.
\bibitem{cediss} 
\Name{Ellenberger C.} 
\Book{Ph.D.~Thesis, Marburg}
\Year{2001}.
\bibitem{bestgen74}
\Name{Bestgen W.} 
\REVIEW{Z. Phys.}{269}{1974}{73}.
\bibitem{bestgenrothen} 
\Name{Bestgen W.\and Rothen F.} 
\REVIEW{Low.\ T.\ Phys.}{77}{1989}{257}.

\end{thebibliography}
\end{document}